\documentclass[cits]{PoS}
\usepackage[intlimits]{amsmath}
\usepackage{subfigure}

\title{On propagators and three-point functions in Landau gauge QCD and
QCD-like theories}

\ShortTitle{On propagators and three-point functions}

\author{\speaker{Reinhard Alkofer},\\
       Institute of Physics, University of Graz,
       Universit\"atsplatz 5,
       A-8010 Graz, Austria\\
       E-mail: \email{reinhard.alkofer@uni-graz.at}}

\author{{Gernot Eichmann$^*$}, \\
       Institute of  Theoretical Physics, Justus-Liebig
       University of Gie\ss en,
       Heinrich-Buff-Ring 16, 35392 Gie\ss en, Germany \\
       E-mail: \email{Gernot.Eichmann@theo.physik.uni-giessen.de}}

\author{{Christian S.\ Fischer$^*$}, \\
       Institute of  Theoretical Physics, Justus-Liebig
       University of Gie\ss en,
       Heinrich-Buff-Ring 16, 35392 Gie\ss en, Germany \\
       E-mail: \email{Christian.Fischer@theo.physik.uni-giessen.de}}

\author{Markus Hopfer, \\
       Institute of Physics, University of Graz,
       Universit\"atsplatz 5,
       A-8010 Graz, Austria\\
       E-mail: \email{Markus.Hopfer@uni-graz.at}}

\author{Milan Vujinovic,\\
       Institute of Physics, University of Graz,
       Universit\"atsplatz 5,
       A-8010 Graz, Austria\\
       E-mail: \email{Milan.Vujinovic@uni-graz.at}}

\author{Richard Williams,\\
       Institute of  Theoretical Physics, Justus-Liebig
       University of Gie\ss en,
       Heinrich-Buff-Ring 16, 35392 Gie\ss en, Germany \\
       E-mail: \email{Richard.Williams@theo.physik.uni-giessen.de}}

\author{ Andreas Windisch\\
       Institute of Physics, University of Graz,
       Universit\"atsplatz 5,
       A-8010 Graz, Austria\\
       E-mail: \email{Andreas.Windisch@uni-graz.at}}

\abstract{Recent progress in our studies of propagators and three-point functions in Landau gauge for
QCD and QCD-like theories is presented. Special emphasis is put on the properties of the
three-gluon vertex and the quark-gluon vertex. The effect of unquenching is investigated.
Furthermore, an exploratory study for a large number of light flavours is described, from where clear
evidence for the qualitative behaviour of propagators in the so-called conformal window can be
extracted.
}

\FullConference{From quarks and gluons to hadronic matter: A bridge too far?\\
		 2-6 September, 2013\\
European Centre for Theoretical Studies in Nuclear Physics and Related Areas (ECT*),
Villazzano, Trento (Italy)}

\begin{document}

\section{Introduction}

In the last decades it has become evident that quantum gauge field theories possess an extremely rich
structure. Taken together with recent developments, most noticeable the discovery of the Higgs
particle and the Planck satellite's very precise observation of the cosmic microwave background, the
current description of physics, based on the Standard Model (SM) of Particle Physics and General
Relativity (GR), has proven to be surprisingly efficacious. With the SM being a collection of three gauge
theories, and GR being diffeomorphism invariant (which can be also understood as a kind of a gauge
invariance), the question arises: ``Why gauge?''

This is exactly the title of Ref.~\cite{Rovelli:2013fga}. In that short article it is
discussed why understanding gauge only as a mathematical redundancy overlooks an important aspect of
physics. Gauge theories describe physical objects relative to each other, in a way similar to the
diffeomorphism invariance inherent to GR.
Nevertheless, interactions in gauge theories are given by gauge-invariant couplings between gauge-dependent
quantities. These ``gauge interactions describe the world because Nature is described by relative
quantities that refer to more than one object''~\cite{Rovelli:2013fga}.

Taking this point of view as given, it entails that a gauge-invariant calculation of
gauge-invariant observables will provide us numbers but no insight into the physical mechanisms
underlying the phenomena provided by a gauge theory. Quarks and gluons as gauge-dependent fields are
unphysical degrees of freedom; nevertheless, they are necessary ingredients for an understanding of
the rich structure of quantum gauge field theories in general and QCD in particular.

The gauge-fixed formulation (analogous to the formulation of GR in a coordinate system) possesses still a
symmetry reflecting the gauge symmetry, the so-called BRST symmetry. For several gauges, it is now known
how to construct the Hilbert space of physical states from a
cohomology of the operator related to the Noether charge of BRST symmetry (see, {\it e.g.},
\cite{Schaden:2013ffa} and references therein). Despite this recent progress,
of these gauges the covariant Landau gauge is still the one which is best understood and therefore
the preferred choice for theoretical as well as practical reasons.

Depending on the number of dimensions and the matter content, quantum gauge field theories can be
asymptotically free, asymptotically safe or renormalization-group trivial. The existence of a
so-called conformal window, in which the gauge theory shows for a restricted interval of the number of
matter flavours a (near) conformal behaviour, is generally accepted but not yet understood.
Related is the question of a Coulomb versus a confining/Higgs phase of the gauge theory.
One also has to note that there is no gauge-invariant order parameter separating the confining from a
Higgs-type realization of gauge symmetries, see, {\it e.g.}, \cite{Caudy:2007sf} and references
therein, thus substantially obscuring the distinction between obviously distinct physical phenomena.

This is not the only reason why understanding the phenomenon of confinement from QCD has proven
to be a truly hard problem. The Wilson loop provides a clear criterion only in the
absence of dynamical quarks; hence, even an unambiguous theoretical definition of confinement is not generally
agreed upon. An operational definition of the confinement phase in the presence of fundamental charges
can be given by requiring two properties: (i)  an unbroken global colour charge (as in the Coulomb
phase) {\bf and} (ii) a mass gap (as in the Higgs phase).

As confinement requires the generation of a mass gap, the question arises whether it is
related to dynamical chiral symmetry breaking (D$\chi$SB) and the U$_A$(1) anomaly, and if so, how? To
this end one notes that D$\chi$SB and the U$_A$(1) anomaly imply the existence of topologically
non-trivial gluon field configurations and quark would-be zero modes as well as the existence of a
supercritical coupling. These phenomena then result in the generation of ``constituent quark masses''
and chiral-symmetry-violating quark-gluon couplings.

All these considerations motivate to extend present studies of QCD's two-point correlators also to
vertex functions such as the quark-gluon, three-gluon and four-gluon vertices. It
should be emphasized that having these Green functions at hand will not only provide some keys to
understand fundamental physics but will also allow to calculate hadronic properties. In this sense, one
bridge from quarks and gluons to hadronic matter is actively under construction. In the following we
want to provide an impression how much progress is being made in closing this gap. 

\section{Green functions in QCD}

In QCD (and similarly QCD-like) gauge theories one has to deal with seven primitively divergent Green functions:
the gluon, ghost, and quark propagators as well as the three-gluon, four-gluon, ghost-gluon and
quark-gluon vertex functions.
Of these, the propagators have received the greatest amount of attention both within functional
approaches~\cite{Pawlowski:2003hq,Alkofer:2004it,Aguilar:2008xm,Fischer:2009tn} and lattice QCD~\cite{Maas:2006qw,Cucchieri:2009zt,Sternbeck:2013zja,Cucchieri:2013nja}, with the vertex functions only being
recently tackled~\cite{Skullerud:2002ge,Alkofer:2008tt,Cucchieri:2008qm,Kellermann:2008iw,Huber:2012kd,Rojas:2013tza,Ahmadiniaz:2013rla} and further investigation needed in most cases. Results for the ghost-gluon
vertex are the most robust, owing to its relative simplicity, and confirm that its dressing is of
mild quantitative importance; thus it is quite safe to
neglect deviations of this vertex from its tree-level form in most studies. In contrast, extant studies of the four-gluon vertex show
enhancement at least in the infrared~\cite{Kellermann:2008iw}; we later use these results in our vertex studies.
Curiously, despite its ubiquity in many calculations, the three-gluon vertex has only recently been
the focus of intense study~\cite{Aguilar:2013vaa,Pelaez:2013cpa,Blum:2014gna,Williams:2014}, building upon the early works of Refs.~\cite{Kim:1979ep,Ball:1980ax}. 
In the next section we will highlight its importance and present an update as to the status of its investigation.

\begin{figure}[t]
\begin{center}
\includegraphics[width=1.01\textwidth]{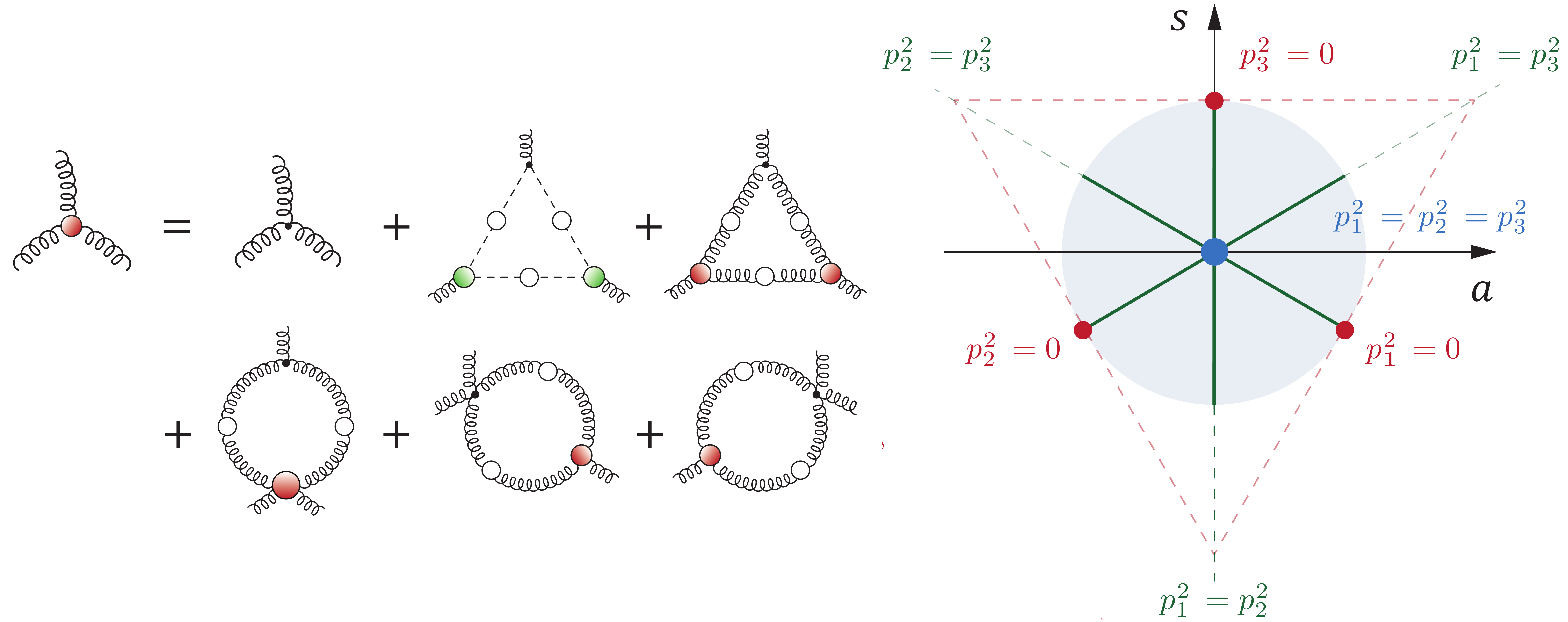} 
\caption{
\label{fig:3gDSE}
\textit{Left panel:} the truncated Dyson-Schwinger equation for the three-gluon vertex function. Filled circles
denote fully dressed vertex functions. In the ghost triangle (2nd diagram on r.h.s.) the ghost-gluon
vertex functions are approximated as bare. In the other diagrams (gluon triangle and swordfish
diagrams) the three-gluon vertex is determined self-consistently. The four-gluon vertex is modelled
with an enhancement against its tree-level structure for intermediate and small momenta.
\textit{Right panel:} phase space for the three-gluon vertex in the $(a,s)$ plane at a slice of fixed $S_0$.}
\end{center}
\end{figure}

\subsection{Three-gluon vertex}

Since the three-gluon vertex epitomises the non-Abelian character of QCD, its degree of influence
on the other Green's functions must be understood. Although the quark-gluon vertex function is of particular
importance since it directly connects the matter to the gauge sector, it has an explicit dependence upon the
details of the gluon self-interaction, as does the gluon propagator~\cite{Alkofer:2008tt,Fischer:2008wy}.
It is anticipated that
the results of these studies will be applied to: computations of the quark-gluon vertex;
unquenched gluon calculations; investigations of the conformal window; the construction of symmetry-preserving
Bethe-Salpeter kernels; and as input for irreducible three-body forces in the covariant Faddeev equation for 
baryons~\cite{Eichmann:2009qa,Eichmann:2011vu}.

\begin{figure}[t]
\begin{center}
\includegraphics[width=0.70\textwidth]{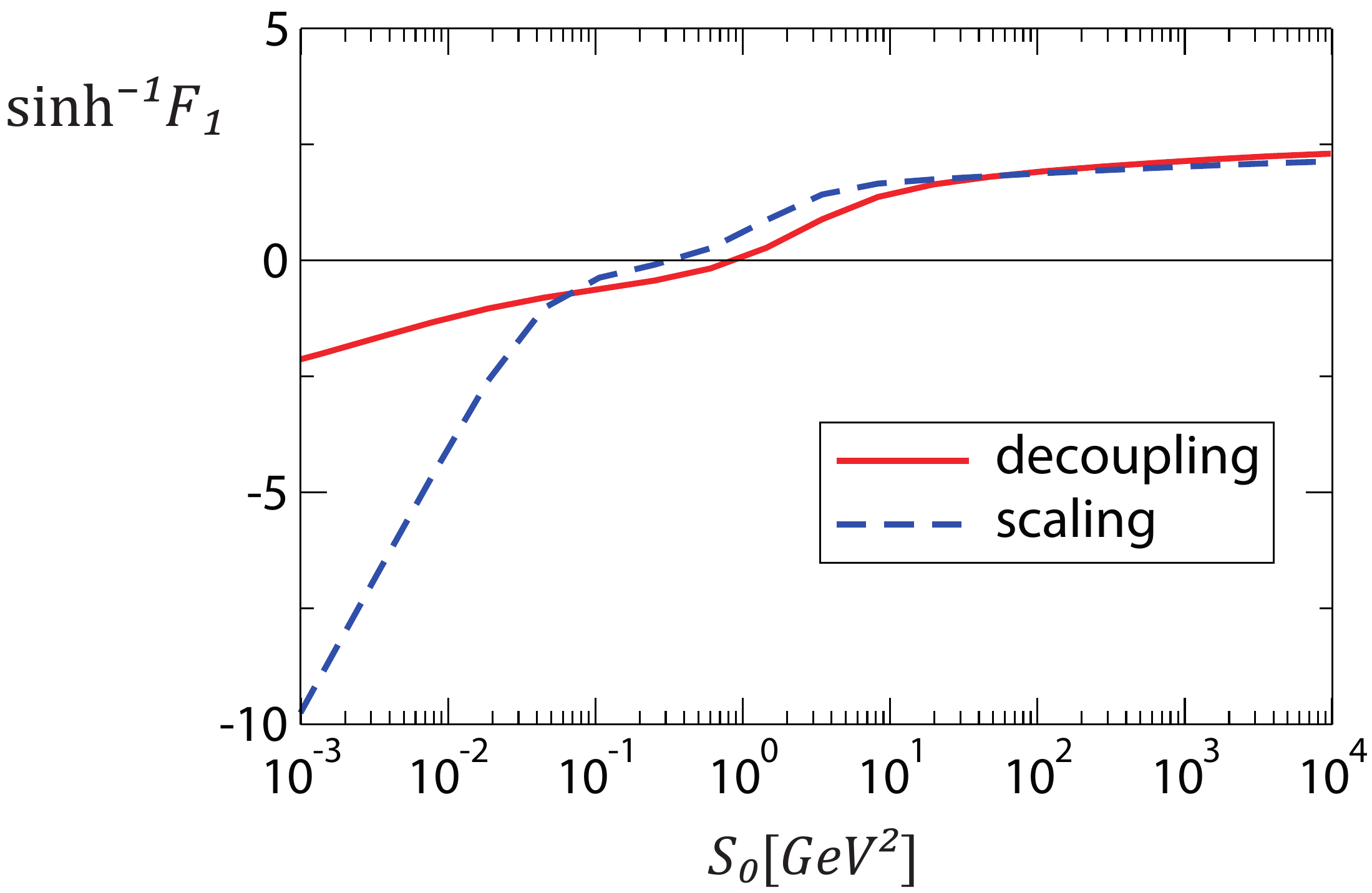} 
\caption{
\label{fig:3gres}
Dressing function $F_1(S_0,a=0,s=0)$ of the tree-level tensor structure in the three-gluon vertex for the scaling 
case and one chosen decoupling solution.}
\end{center}
\end{figure}


In Ref.~\cite{Williams:2014} the truncated Dyson-Schwinger equation (DSE) displayed in Fig.~\ref{fig:3gDSE} 
has been solved in both scaling and decoupling scenarios.
The truncation is chosen such that all ultraviolet and infrared dominant terms
are retained, {\it cf.} Ref.~\cite{Alkofer:2008dt}, at the expense of omitting two-loop diagrams and
vertices without a tree-level counterpart. An additional cyclic permutation ensures
Bose symmetry of the vertex with respect to its external legs.
Ultimately one is interested in inserting the three-gluon vertex in other functional equations and/or 
hadronic matrix elements. 
The transversality of the gluon in Landau gauge entails that it is sufficient to calculate
only the transversely projected part of the vertex since it already carries the full dynamics.
This decreases the number of independent Lorentz tensor structures from fourteen to four 
which are all retained in the calculation. Further emphasis in Ref.~\cite{Williams:2014} has been placed on exploring
the consequences of Bose symmetry, both on the phase space of Lorentz invariants and the tensor basis.
The `natural' set of momentum invariants as inferred from the permutation group is
        \begin{equation}
             S_0 = \frac{1}{6}\,(p_1^2+p_2^2+p_3^2)\,, \quad
             a = \sqrt{3}\,\frac{p_2^2-p_1^2}{p_1^2+p_2^2+p_3^2}\,,  \quad
             s = \frac{p_1^2+p_2^2-2p_3^2}{p_1^2+p_2^2+p_3^2}\,,
        \end{equation}
where $S_0$ is totally symmetric and $(a,s)$ form a doublet of mixed symmetry.
The phase space in $(a,s)$ for a fixed value of $S_0$ is illustrated by the filled circle in the right panel of Fig.~\ref{fig:3gDSE}.
Various special momentum configurations such as the symmetric limit $p_1^2=p_2^2=p_3^2$ or the three soft kinematic points $p_i^2=0$
are also shown. The latter are interesting because they can exhibit soft singularities in the vertex and thereby have an impact
on the stability of the DSE iteration.

In Fig.~\ref{fig:3gres} the numerical result for the leading tree-level vertex dressing function is shown as a function 
of $S_0$ in the symmetric configuration $a=s=0$.
Its qualitative behaviour is already determined by the first two
terms in Fig.~\ref{fig:3gDSE}, the tree-level term plus the ghost triangle.
This is due to a partial cancellation of the remaining terms which are individually large,
namely the gluon triangle and the swordfish diagrams. However, their inclusion still has a quantitative impact on the result.
The ghost loop is negative and exceeds in absolute value the tree-level term for very small momenta, thereby producing a zero crossing
which is a robust feature in both scaling and decoupling solutions~\cite{Blum:2014gna,Williams:2014}. 
In contrast to the power-law behaviour of the three-gluon vertex
in the deep infrared for the scaling scenario, in decoupling it is instead log-divergent~\cite{Aguilar:2013vaa}.

The presence of a zero crossing in the three-gluon vertex in four dimensions is surprising, but not forbidden. Such a feature has
been seen on the Lattice in 2 and 3 dimensions~\cite{Cucchieri:2008qm}; the four-dimensional case is not so clear at present. The
phenomenological relevance of such a negative three-gluon vertex is a topic for future studies, where it is presumably important
for exotic/hybrid mesons in addition to three-body interactions in baryons. An understanding of this vertex is also required for
QCD-like theories such as technicolor wherein the gauge-sector and its coupling to matter fields may be quite different to what
we have come to accept in QCD.

\subsection{Quark-gluon vertex}

In studies of the strong interaction, our desire to understand the wealth of hadronic properties requires connection
to be made between the fundamental, but unphysical, quarks and gluons. Hence, in such studies the quark-gluon vertex
is of prime importance because it triggers D$\chi$SB in the effective quark-gluon 
interaction~\cite{Skullerud:2003qu,Maris:2003vk,Bhagwat:2004hn,Matevosyan:2006bk,Alkofer:2008et,Fischer:2009jm,Chang:2009zb,Bashir:2012fs,Chang:2013nia}.
Other non-perturbative effects such as pion-cloud corrections are also contained within that 
vertex~\cite{Watson:2004jq,Thomas:2008bd,Fischer:2007ze,Fischer:2008wy,Heupel:2013zka}, 
together with the dominant dependence on quark flavour necessary to make connection with the 
heavy quark limit~\cite{Maris:2006ea,Popovici:2010mb,Blank:2011ha}.

\begin{figure}[t]
\begin{center}
\includegraphics[width=0.45\textwidth]{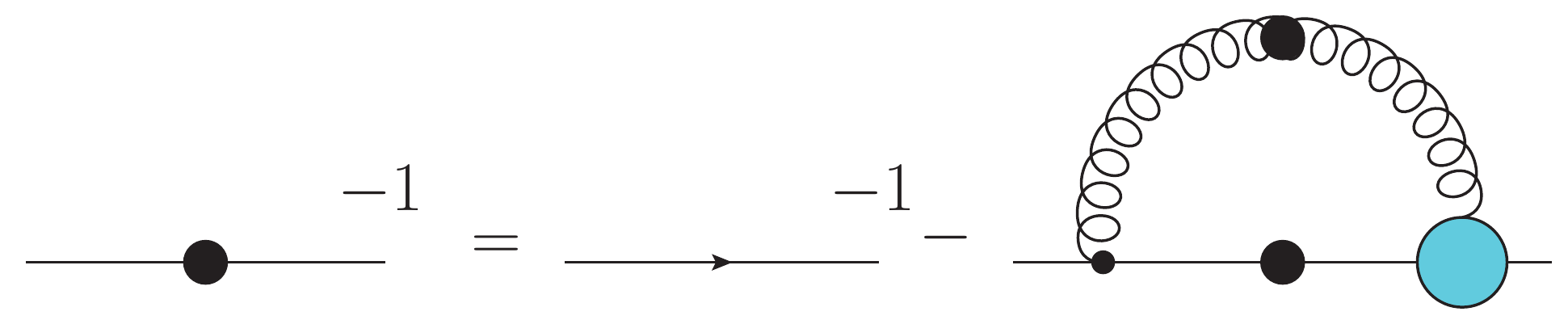}\hfill
\includegraphics[width=0.50\textwidth]{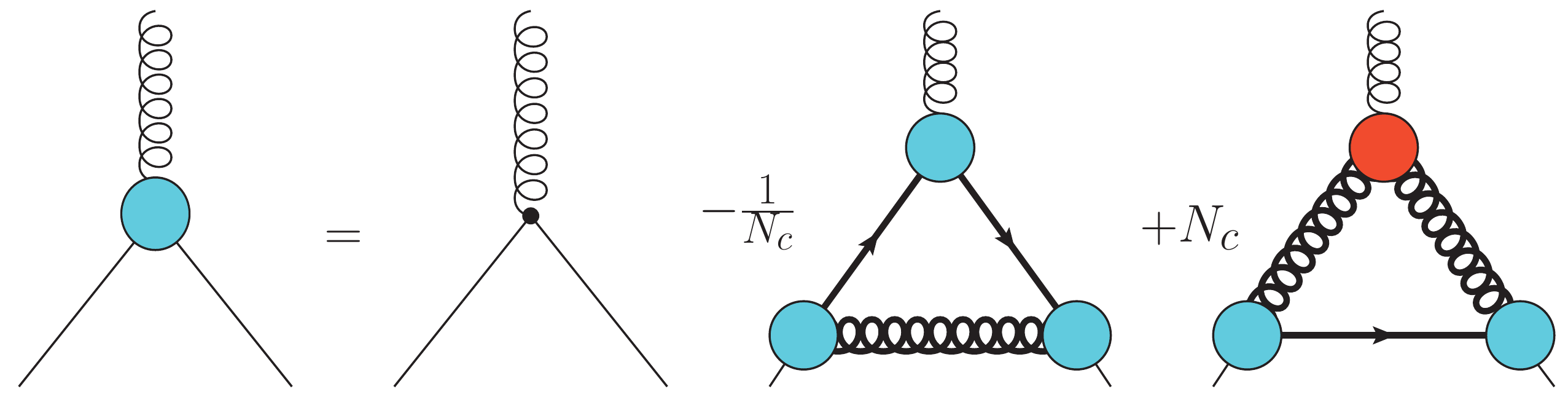} 
\end{center}
\caption{The coupled system for the quark propagator and truncated quark-gluon vertex  Dyson-Schwinger
equations. The Abelian diagram, featuring three dressed quark-gluon vertices, is $N_c^2$ suppressed with
respect to the non-Abelian diagram that contains the gluon self-interaction.
\label{fig:QGVdse}}
\end{figure}

Based on earlier investigations~\cite{Alkofer:2008tt,Fischer:2009jm}, the quark-gluon vertex function
is currently calculated~\cite{Windisch:2014,Williams:2014xxx} in a truncation employing the (quenched) gluon propagator
and a model for the three-gluon vertex, see Fig.~\ref{fig:QGVdse}.
The truncation is based on arguments from $n$-particle-irreducible actions in accordance with an infrared
analysis of the scaling solution \cite{Alkofer:2008tt}.\footnote{It should be noted that a
scaling-type quark-gluon vertex provides a possible explanation of the U$_A$(1) anomaly within
functional approaches \cite{Alkofer:2008et}.}
Due to technical complications we present results with the Abelian diagram in Fig.~\ref{fig:QGVdse} omitted.
The remaining non-Abelian diagram is solved in accordance to \cite{Hopfer:2013np} and, independently,
by folding the quark-gluon four-point function, which is analogous 
to the quark-Compton vertex~\cite{Eichmann:2012mp},
together with the three-gluon vertex~\cite{Williams:2014}.
For convenience we employ different basis decompositions internally and externally 
and perform rotations in each iteration step.

\begin{figure}[!ht]
\begin{center}
\subfigure{\includegraphics[width=0.45\textwidth]{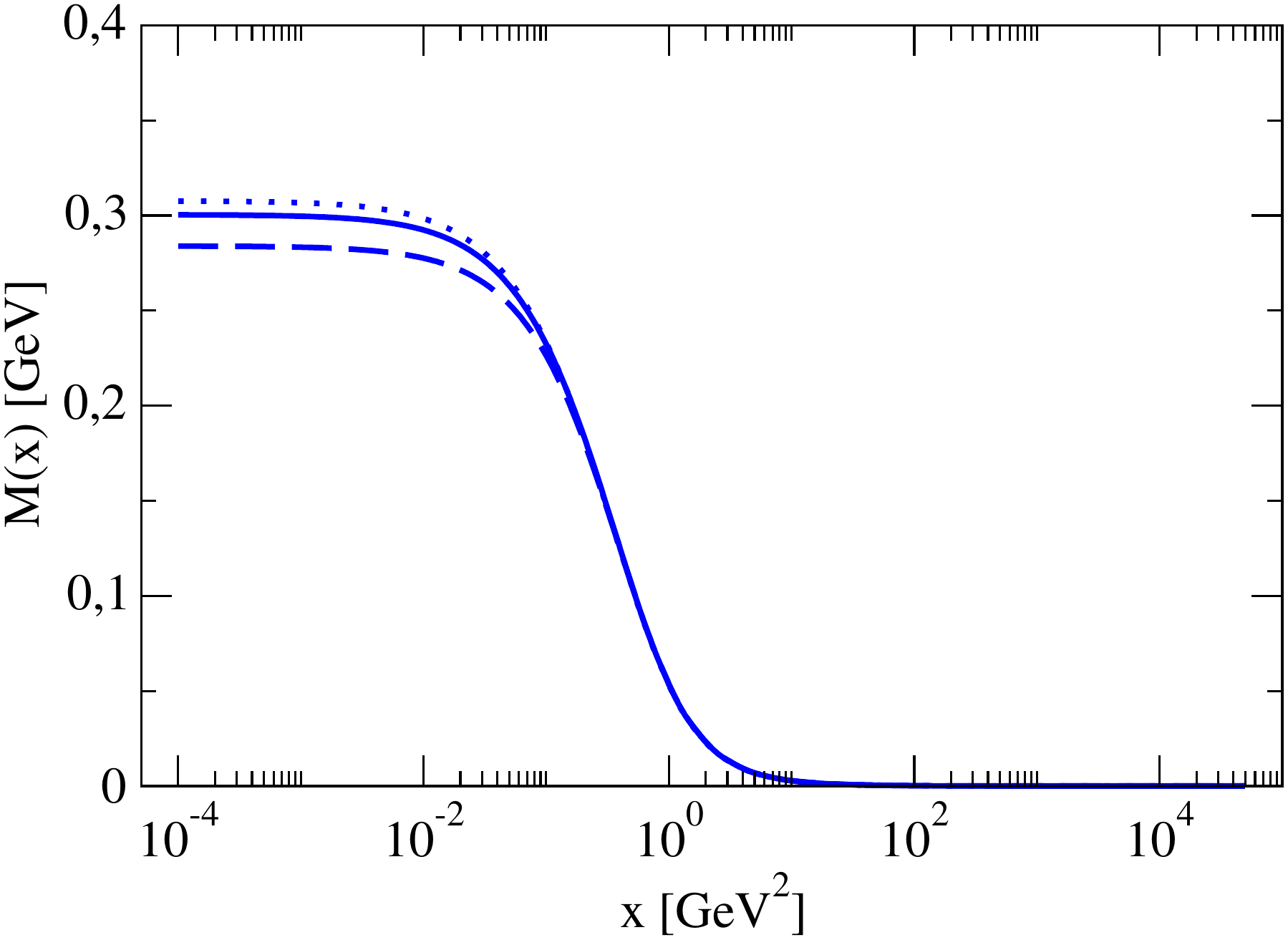}\label{fig:qgv_DynamicalMass}}\hspace{0.3cm}
\subfigure{\includegraphics[width=0.5\textwidth]{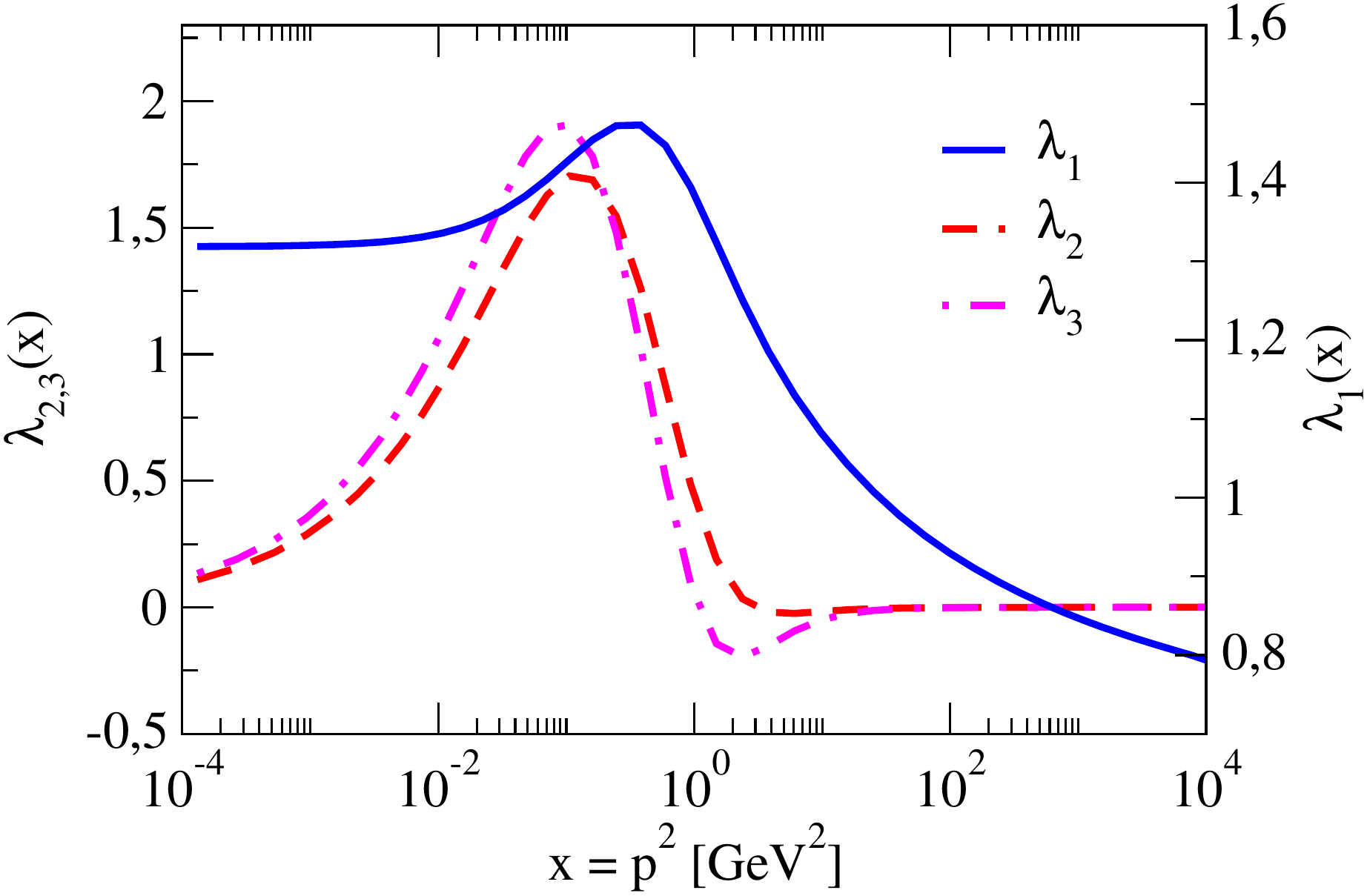}\label{fig:QGVresults}}
\end{center}
\caption{Left diagram: The quark mass function $M(p^2)$ using different models for the three-gluon vertex.
Right diagram: Results for some of the coefficient functions of the quark-gluon vertex at symmetric momenta
$x=p_1^2=p_2^2=p_3^2$.}
\end{figure}

One can infer from Fig.~\ref{fig:QGVresults} that
the tree-level tensor structure, parametrised by the coefficient function $\lambda_1$, is
significantly enhanced below the $1-2$~GeV scale.
$\lambda_3$ is an example of a tensor structure which would be absent if chiral
symmetry were realized in the Wigner-Weyl mode; it is generated self-consistently and non-perturbatively through D$\chi$SB.
Our preliminary results indicate that the tensor structures which are even in the number of Dirac matrices 
are of similar size to those with an odd number of Dirac matrices; in particular see
$\lambda_2$ vs. $\lambda_3$ in Fig.~\ref{fig:QGVresults}. Fig.~\ref{fig:qgv_DynamicalMass} shows the quark mass function.

\subsection{Coupling quarks and gluons: unquenching effects}

\begin{figure}[t]
\begin{center}
\includegraphics[width=0.99\textwidth]{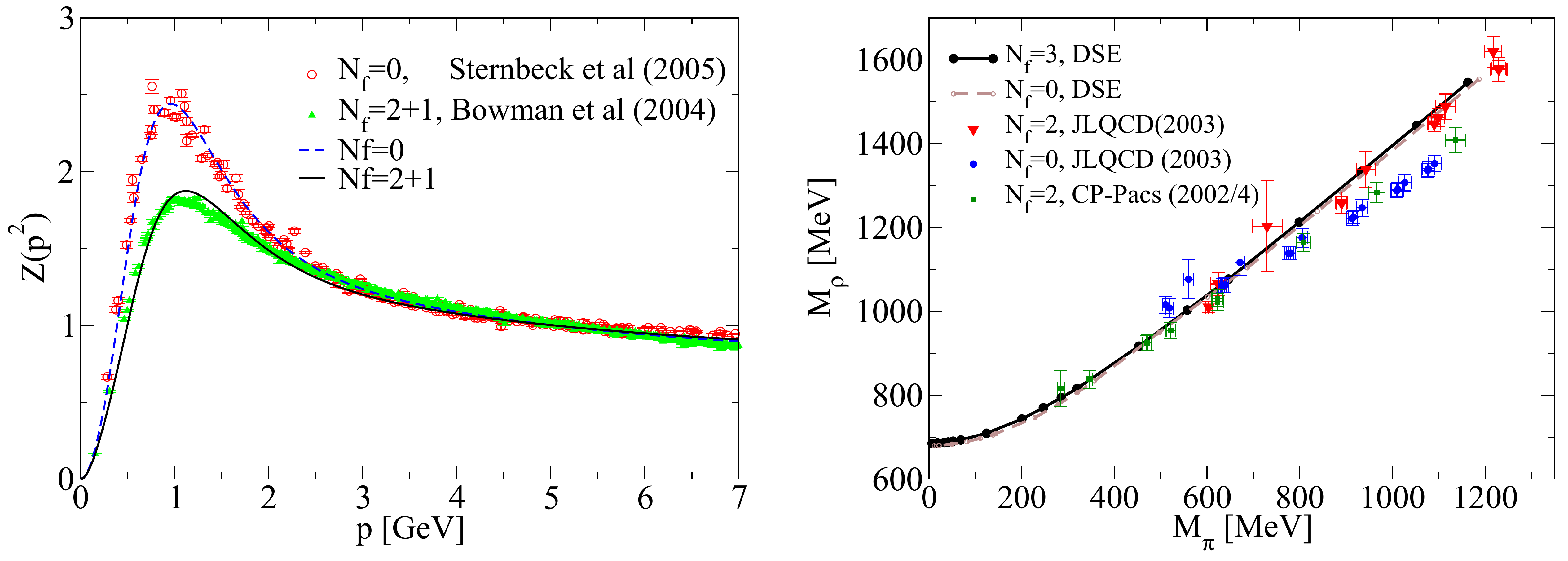}
\caption{
\label{fig:glu_unq}
Left diagram: Quenched \cite{Huber:2012kd} and unquenched ($N_f=2+1$) gluon dressing
function from DSEs compared with the lattice results of
\cite{Sternbeck:2005tk,Bowman:2004jm}.
Right diagram:
quenched and unquenched results for the vector meson mass (figure adapted from
Ref.~\cite{Fischer:2005en}) compared to lattice results \cite{Namekawa:2004bi,AliKhan:2001tx,Aoki:2002uc}.}
\end{center}
\end{figure}
\begin{figure}[t]
\begin{center}
\includegraphics[width=0.98\textwidth]{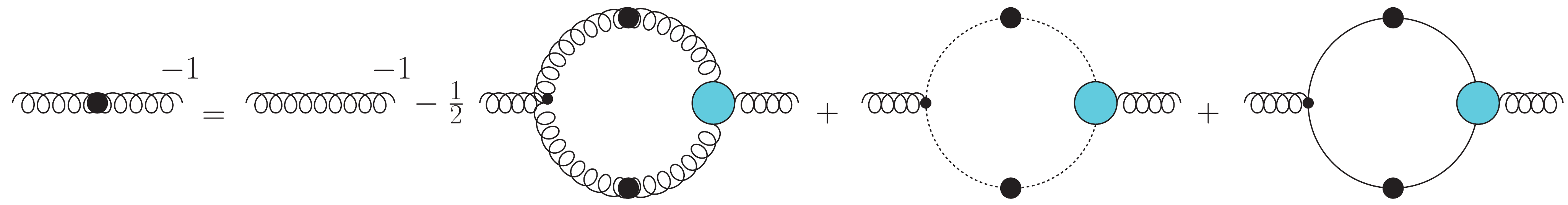}\\[3mm]
\includegraphics[width=0.45\textwidth]{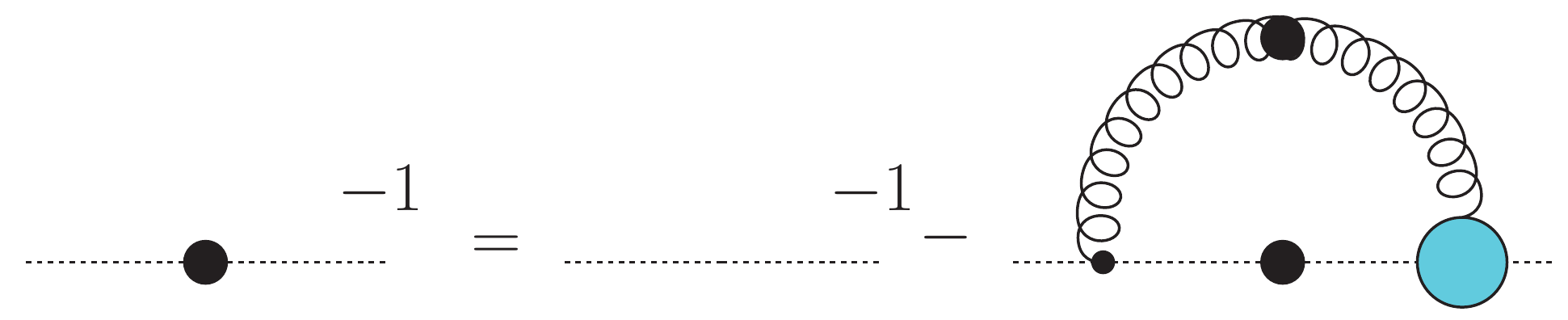} \hfill
\includegraphics[width=0.45\textwidth]{QuarkDSE.pdf}
\caption{
\label{fig:DSEs}
Dyson-Schwinger equations for the gluon, ghost and quark propagators (the gluon one is truncated to
the one-loop level).}
\end{center}
\end{figure}

Unquenching effects in QCD are associated with closed quark loops
on the diagrammatic level. These appear directly on the level of
the propagator DSEs of QCD only in the equation for the gluon
propagator, as displayed in Fig.~\ref{fig:DSEs}. In contrast to
the ghost and gluon contributions to the system, the quark loop
has a screening effect, reflecting its positive contribution to the
ghost and gluon's anomalous dimensions $\delta$ and $\gamma$ governing
the logarithmic running at large momenta,
\begin{equation}
\delta = \frac{-9 N_c}{44 N_c - 8 N_f}\,, \hspace*{10mm}
\gamma = \frac{-13 N_c + 4 N_f}{22 N_c - 4 N_f}\,.
\end{equation}
The resulting modification of the gluon and ghost propagators in the
nonperturbative mid-momentum regime has been studied first in
Ref.~\cite{Fischer:2003rp} within the Dyson-Schwinger framework for
a scaling type of solution. For decoupling, a corresponding study is
shown in the left diagram of Fig.~\ref{fig:glu_unq}. Here we use the
quenched approximation scheme of Ref.~\cite{Huber:2012kd} as a
starting point and back-couple $N_f=2+1$ quark flavours to the gluon
system using a vertex construction as described in \cite{Fischer:2003rp}
(for results on unquenching in the background gauge formalism
see \cite{Aguilar:2013hoa}). The resulting unquenching effect affects
mainly the mid-momentum region, and is of the order of 25 percent,
agreeing nicely with the effects seen in the lattice data of
Refs.~\cite{Sternbeck:2005tk,Bowman:2004jm} (for recent results see also
\cite{Ayala:2012pb}). All curves have been normalized to one at $p=5$ GeV.
The corresponding unquenching effects in the quark propagator are somewhat
smaller and have been studied in Ref.~\cite{Fischer:2003rp}. Also an extension to the
Bethe-Salpeter framework of light mesons is available and produced results
for the dependence of the vector meson mass on the underlying current
quark mass \cite{Fischer:2005en}, shown in the right diagram of
Fig.~\ref{fig:glu_unq}.

Unquenching effects, however, do not occur only on the level of the
propagators. In the Dyson-Schwinger equation for the quark-gluon vertex
unquenching effects can be effectively summed up into contributions from
the backreaction of mesons onto the quark-gluon system. The corresponding
framework has been formulated in \cite{Fischer:2007ze} and applied to the light
meson spectrum in \cite{Fischer:2008wy}. It turns out that the unquenching
effects on meson masses due
to the pion backreaction are dominant as compared to the above discussed
gluon polarization effects. This is so because the pion backreaction effects
are directly manifest in the kernel of the Bethe-Salpeter equations, whereas
the gluon polarization effects show up only indirectly. In contrast, when it
comes to the Yang-Mills sector of the system, the order of effects may very
well be expected to be reversed. This is the underlying rationale for
our treatment of large-$N_f$ QCD detailed in the next subsection.

\section{Green functions in QCD-like theories}

\begin{figure}
\centering
\includegraphics[width = 0.38\textwidth]{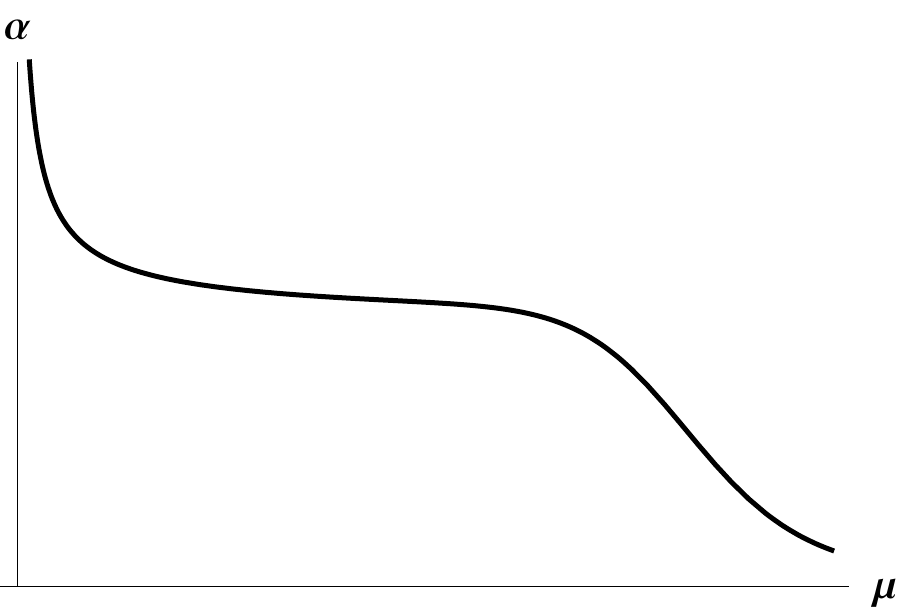}
\includegraphics[width = 0.38\textwidth]{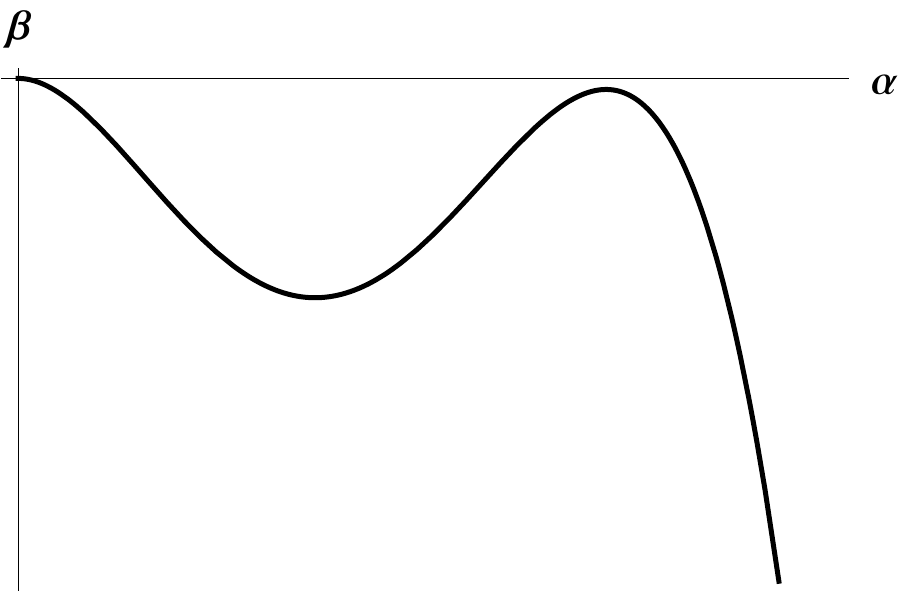}
\caption{The scale dependence of gauge coupling (left) and the corresponding $\beta$ function (right) in a theory with nearly 
conformal dynamics.}
\label{fig:TC}
\end{figure}

Inspired by QCD, Technicolor (TC) as posited by Weinberg and Susskind~\cite{Weinberg:1979bn,Susskind:1978ms} describes the 
Higgs boson as a composite particle of yet undiscovered constituents called techniquarks. 
In its original incarnation, a scaled-up version of QCD, TC was found to have many problems when faced with experimental data. 
Today, it is considered highly desirable (although not necessary) for a TC theory to possess nearly conformal (or 'walking') 
dynamics where the gauge coupling of the theory very slowly evolves over a wide range of energies, as sketched in Fig.~\ref{fig:TC}.
Such a theory would be able to provide masses for the heaviest of Standard Model quarks without giving too much strength to flavour-changing 
neutral currents~\cite{Holdom:1981rm}. Moreover, theories which feature walking dynamics are expected to have unusually light (by QCD standards) 
technihadrons~\cite{Dietrich:2005jn}. Without this property it would be very hard to obtain a composite Higgs boson with a mass of just $125$~GeV, 
even if one takes into account "radiative corrections to the Higgs mass", the corrections to Higgs mass coming from top-quark loops~\cite{Foadi:2012bb}.

Whether a theory displays (near) conformal behaviour depends on the number of colours 
and techniquark flavours; it also depends on the dimensions 
of the representation in which the matter field belong. Various TC scenarios have been
investigated in both the continuum and on the lattice~\cite{Hasenfratz:2011xn,Fodor:2012ty}, 
focusing mainly on the nature of the infrared fixed point and on 
physical observables such as technihadron masses and decay constants. Thus far,
few studies in the context of Green's functions have been made~\cite{August:2013jia}. 
In the following we present the results of a Dyson-Schwinger study of Yang-Mills 
propagators in a theory with a large number of fundamentally charged quark flavours.

\subsection{Many light flavours and the conformal window}

The asymptotic freedom of QCD is lost if the number of quark flavours exceeds sixteen. Considering
the number of light flavours $N_f$ as an additional parameter, the question arises whether QCD is
confining and/or chiral-symmetry-breaking up to $N_f=16$. By now it is generally accepted that this
will be not the case. For an interval $N_f^{crit}\le N_f \le 16$ the theory will display a (near)
conformal behaviour. The exact value of $N_f^{crit}$ is not known and
might be somewhere in between six and twelve.
Since the theory displays a drastically different behaviour when tuning up $N_f$,
one anticipates also the Green functions to
undergo significant changes, especially if their infrared behaviour is considered. As a first question
one might ask how the gauge boson and ghost propagator will behave in the conformal window.

Here we present a model calculation \cite{Hopfer:2014} to provide a qualitative answer
to this question. We emphasize that the employed model is not sophisticated enough to give
quantitatively reliable results, neither for $N_f^{crit}$ nor for precise values of the propagator.
However, it is constructed such that the change for the propagators and the resulting running coupling
can be exemplified. To this end, the coupled Dyson-Schwinger equations for the three propagators are
solved self-consistently (the gluon DSE is truncated to the one-loop level) with the three-point
functions being modelled. The ghost-gluon vertex is taken bare by the same arguments as given above.
The three-gluon and the quark-gluon vertex are modelled following
Ref.~\cite{Fischer:2003rp}.\footnote{For the use of a similar model for the quark-gluon vertex see,
{\it e.g.}, \cite{Aguilar:2010cn}.} It turns out that the results are quite insensitive to the
model for the three-gluon vertex whereas a strong sensitivity to the quark-gluon vertex is found. 
The resulting value $N_f^{crit} \approx 4.5$ is certainly unrealistically
small and will increase with an improved model for the quark-gluon vertex~\cite{Hopfer:2014}. 
Phrased otherwise, to gain
quantitative predictions a reliably calculated quark-gluon vertex is needed.
However, as emphasized in this exploratory study we want to focus on the behaviour of the propagators.

\begin{figure}[t]
\center
{\includegraphics[width=0.5\columnwidth]{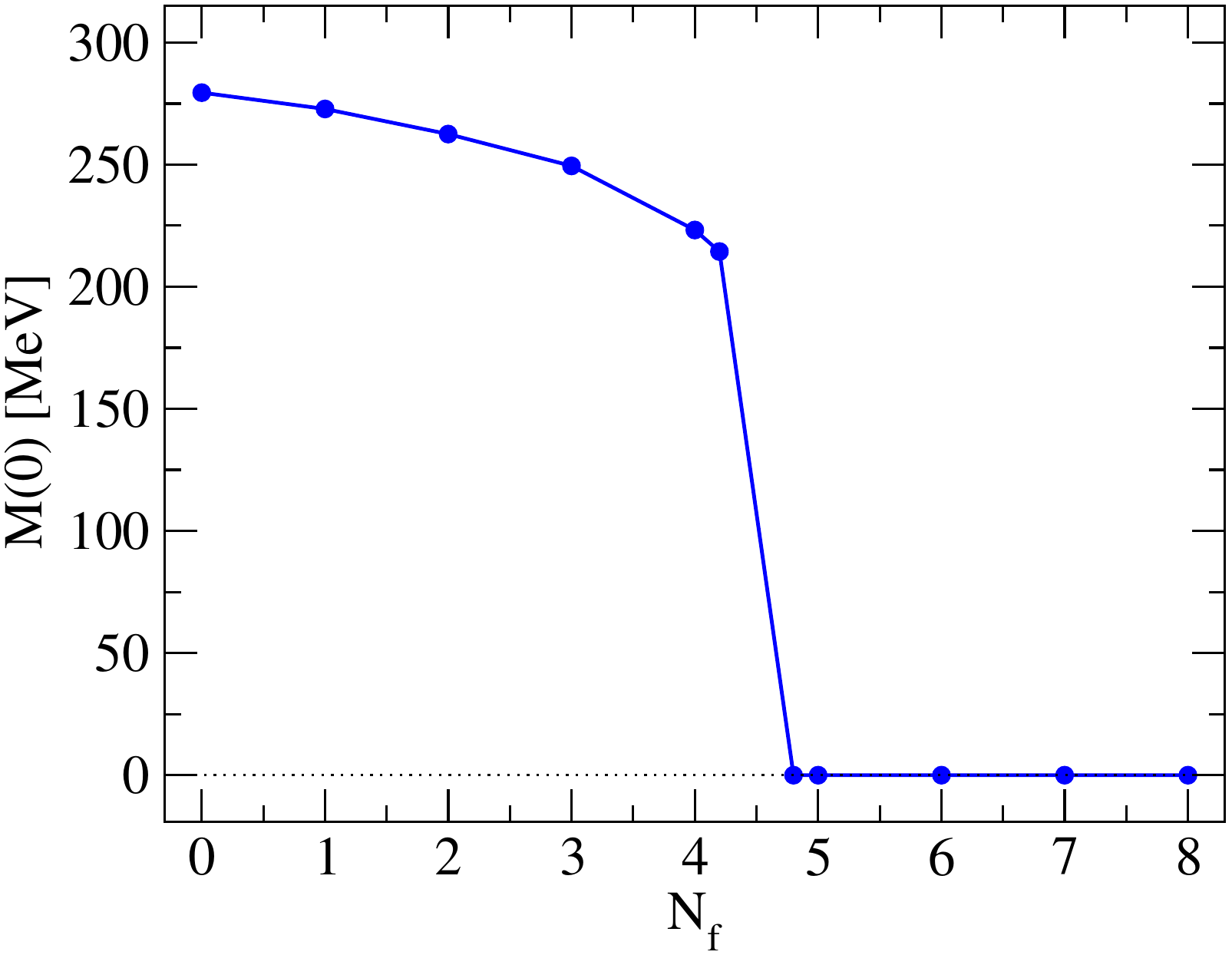}}
\caption{
The quark mass function $M(p^2)$ in the chiral limit and the 
limit of vanishing external momentum $p^2$ for a
different number of light flavours, $N_f$.
Above $N_f^{crit} \approx 4.5$ no dynamical mass is generated. 
(Note that the line is only drawn to guide the eye.)}
\label{fig:flavour}
\end{figure}
As can be seen from Fig.~\ref{fig:flavour}, D$\chi$SB and dynamical quark mass
generation takes only place for $N_f<N_f^{crit}$. 
\begin{figure}[t]
\center
\includegraphics[width=0.48\columnwidth]{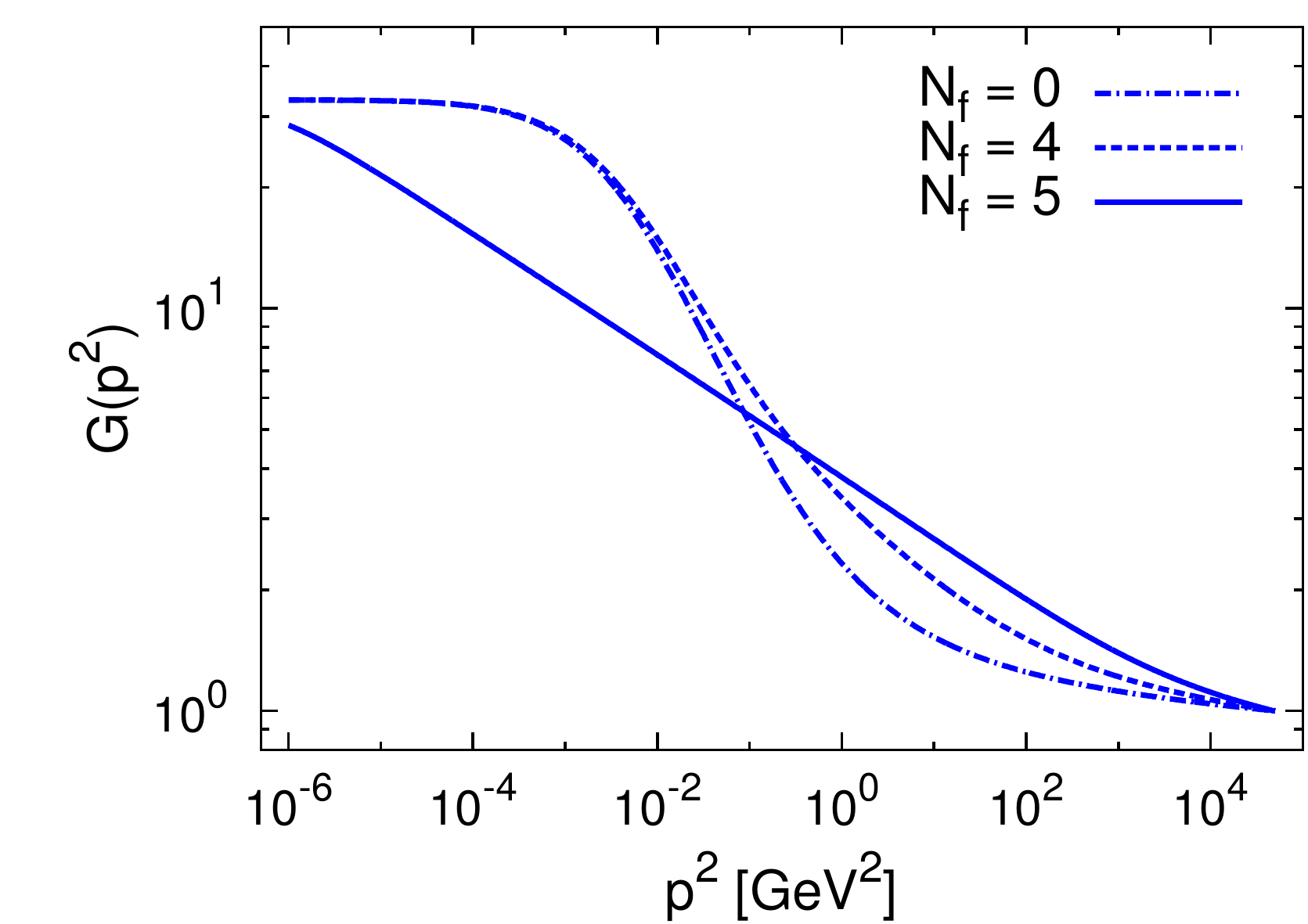}
\includegraphics[width=0.48\columnwidth]{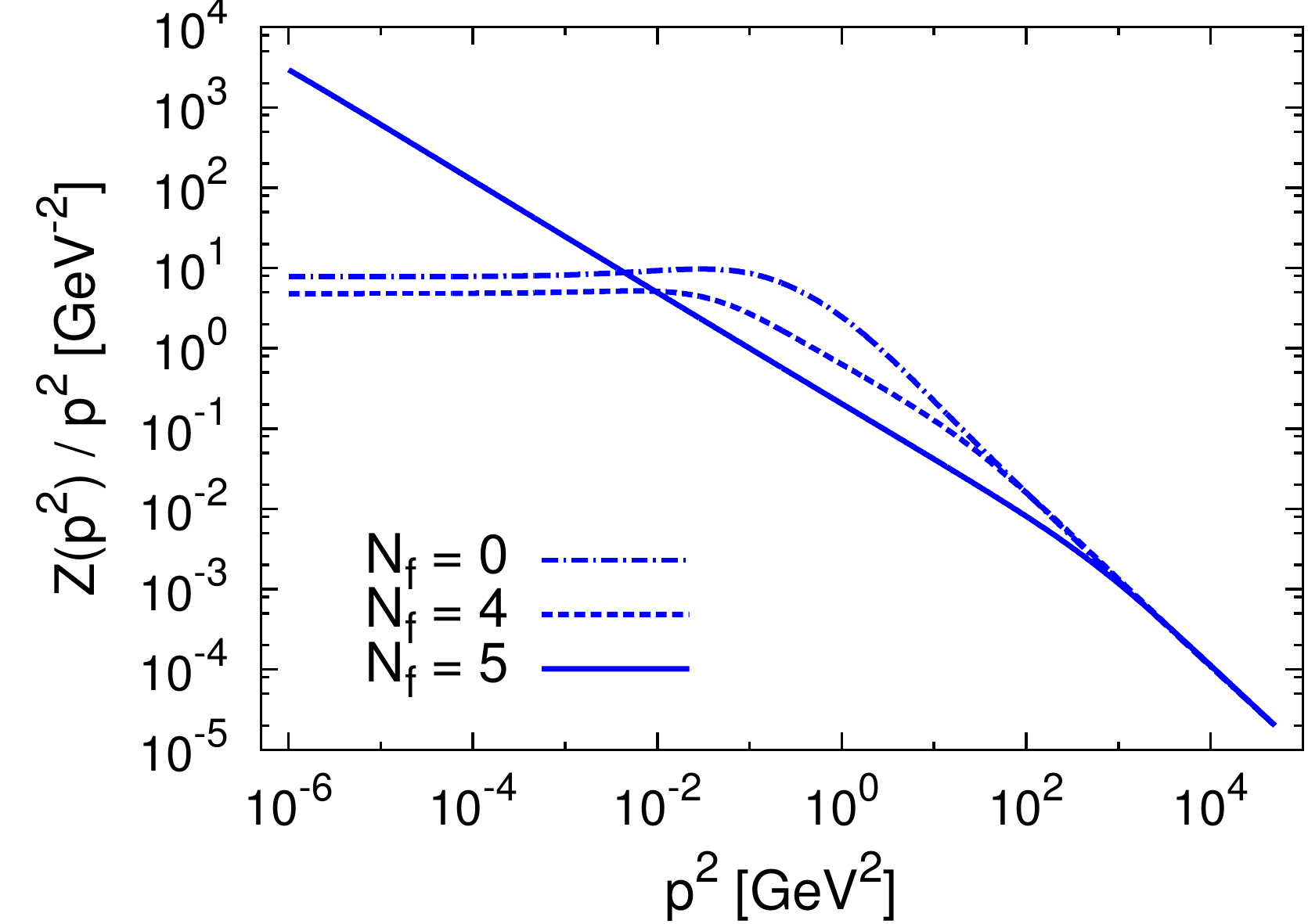}\\
\includegraphics[width=0.48\columnwidth]{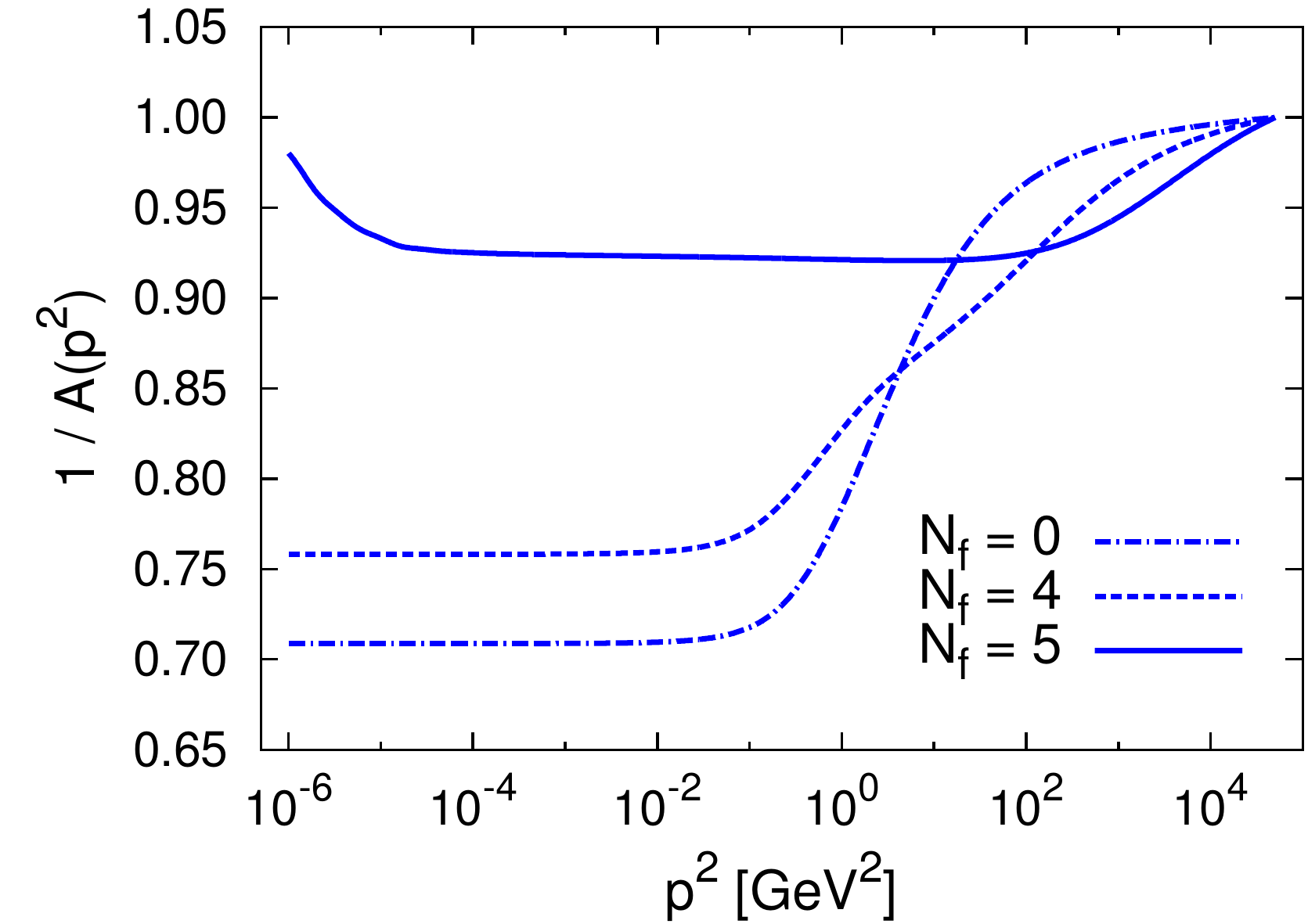}
\includegraphics[width=0.48\columnwidth]{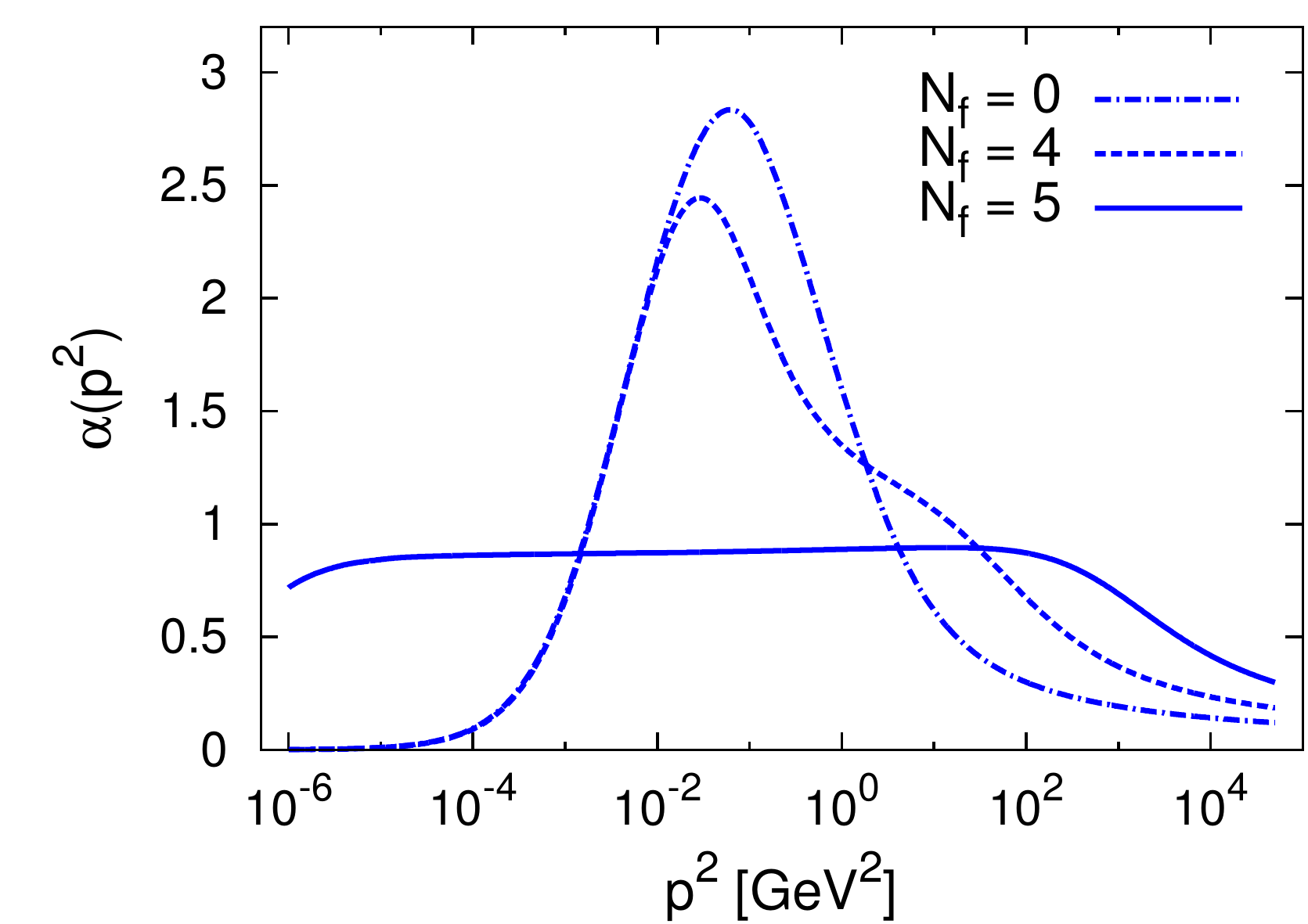}
\caption{Results for  the ghost dressing function $G(p^2)$,
the gluon propagator $Z(p^2)/p^2$,  the quark wave-function renormalization
$A^{-1}(p^2)$  and the running coupling $\alpha(p^2)$
using an ansatz for the quark-gluon vertex.}
\label{fig:1BCxG2_Results}
\end{figure}
In the upper left panel of Fig.~\ref{fig:1BCxG2_Results} the ghost dressing function is displayed. The
anticipated change in its behaviour when increasing $N_f$ above $N_f^{crit}$ is clearly seen. 

The infrared enhanced (although finite) dressing function of the decoupling-type ghost propagator
changes to a power law behaviour for a large region of momenta above $N_f^{crit}$.
Even more interesting is the case of the gluon propagator
(upper right panel): for $N_f \geq N_f^{crit}$ the infrared screened gluon propagator  rises with a power law
towards smaller momenta. The
inverse screening length ``$m(0)$'' even increases slightly with $N_f$ before it drops to zero.
Last but not least, the coupling
displays a plateau characteristic for a ``walking'' coupling,
see the lower right panel of Fig.~\ref{fig:1BCxG2_Results}.

\section{Summary, conclusions, and outlook}

In these proceedings we have presented the results of some recent studies of propagators and
three-point functions in Landau gauge for QCD, and QCD with a hypothetically large number of light
flavours. The most important results are:
\begin{itemize}
\item[-] verification of a zero in the three-gluon vertex.

\item[-] that dynamical breaking of chiral symmetry also takes place in the quark-gluon vertex functions.

\item[-]  a drastic change of gauge boson and ghost propagator for large $N_f$ ({\it i.e.}, conformal window) QCD, wherein
 chiral symmetry is no longer dynamically broken. A plateau reminiscent of ''walking'' is seen in the running coupling.
\end{itemize}
It is fair to say that our understanding of QCD Green's functions has improved and still is substantially
increasing. We will soon see further application of calculated propagators and vertex functions in
QCD bound state equations and thus hadron physics. In other words:  the bridge is likely not too far;
 the community is actively contributing towards its construction.

\bigskip

\noindent
\textbf{Acknowledgments}
We acknowledge support from the Helmholtz International Center for FAIR within the LOEWE
program of the State of Hesse, by BMBF under contract 06GI7121, by the DFG TR-16,
from the Austrian Science Fund (FWF) under project numbers M1333-N16, J3039-N16 and P25121-N27.
MH, MV and AW acknowledge support from the Doktoratskolleg ''Hadrons in Vacuum, Nuclei and Stars`` of the 
Austrian Science Fund, FWF DK W1203-N16.

\providecommand{\href}[2]{#2}\begingroup\raggedright
\endgroup

\end{document}